\documentclass[conference]{IEEEtran}
\usepackage{cite}
\usepackage[utf8]{inputenc}
\usepackage{booktabs}       
\usepackage{amsfonts}       
\usepackage{amsmath}
\usepackage{amssymb}
\usepackage{color, soul}
\usepackage{graphicx}
\graphicspath{ {./figures/} }
\usepackage{subfig}
\usepackage[sort&compress,numbers]{natbib}
\usepackage{hyperref}
\usepackage{multirow}
\usepackage{multicol}
\usepackage{siunitx}

\title{Assessment and manipulation of the computational \\
capacity of \textit{in vitro} neuronal networks \\
through criticality in neuronal avalanches}

\author{\IEEEauthorblockN{Kristine Heiney\IEEEauthorrefmark{1}\IEEEauthorrefmark{2}\IEEEauthorrefmark{4},
Ola Huse Ramstad\IEEEauthorrefmark{3},
Ioanna Sandvig\IEEEauthorrefmark{3},
Axel Sandvig\IEEEauthorrefmark{3}, and
Stefano Nichele\IEEEauthorrefmark{1}}
\IEEEauthorblockA{\IEEEauthorrefmark{1}\textit{Department of Computer Science,
Oslo Metropolitan University, Oslo, Norway}}
\IEEEauthorblockA{\IEEEauthorrefmark{2}\textit{Department of Computer Science,
Norwegian University of Science and Technology,
Trondheim, Norway}}
\IEEEauthorblockA{\IEEEauthorrefmark{3}\textit{Department of Neuromedicine and Movement Science,
Norwegian University of Science and Technology,
Trondheim, Norway}}
\IEEEauthorblockA{\IEEEauthorrefmark{4}Email: kristine.heiney@oslomet.no}}

\begin{document}

\maketitle

\begin{abstract}
In this work, we report the preliminary analysis of the electrophysiological behavior of \textit{in vitro} neuronal networks to identify when the networks are in a critical state based on the size distribution of network-wide avalanches of activity.
The results presented here demonstrate the importance of selecting appropriate parameters in the evaluation of the size distribution and indicate that it is possible to perturb networks showing highly synchronized---or supercritical---behavior into the critical state by increasing the level of inhibition in the network.
The classification of critical versus non-critical networks is valuable in identifying networks that can be expected to perform well on computational tasks, as criticality is widely considered to be the state in which a system is best suited for computation.
This type of analysis is expected to enable the identification of networks that are well-suited for computation and the classification of networks as perturbed or healthy.
This study is part of a larger research project, the overarching aim of which is to develop computational models that are able to reproduce target behaviors observed in \textit{in vitro} neuronal networks.
These models will ultimately be used to aid in the realization of these behaviors in nanomagnet arrays to be used in novel computing hardwares.

\end{abstract}

\section{Introduction} \label{intro}

Current computing technology is based on the von Neumann architecture, in which tasks are performed sequentially and control, processing, and memory are each allocated to structurally distinct components.
With this architecture, conventional computers struggle to cope with the rising demand for data processing and storage.
Furthermore, although recent advancements in machine learning technology have conferred great advantages to our data handling capabilities, processing continues to be performed on conventional hardware that has no inherent learning capabilities and thus requires huge amounts of training data, computational time, and computing power.

To continue to fulfill the rapidly growing computing demands of the modern day, it will be necessary to develop novel physical computing architectures that are scalable, capable of learning, energy-efficient, and fault-tolerant.
The use of self-organizing substrates showing an inherent capacity for information transmission, storage, and modification \cite{langton1990edgeofchaos} would bring computation into the physical domain, enabling improved efficiency through the direct exploitation of material and physical processes for computation \cite{broersma2017computational, konkoli2018reservoir, jensen2018spinice}.
Some key properties of self-organizing systems that make them well-suited for computational tasks include their lack of centralized control and their adaptive response to changes in their environment \cite{heylighen1999}.
Such systems are composed of many autonomous units that interact with each other and the environment through a set of local rules to give rise to organized emergent behaviors at a macroscopic scale.
This type of spontaneous pattern formation is fairly common in nature, and there has been recent interest in determining how to develop interaction rules to generate various desired emergent behaviors \cite{doursat2013morphogenetic}, including those geared toward computation.
In addition, it has been demonstrated that self-organizing substrates, such as magnetic arrays and self-assembling molecules, can be used as computational reservoirs---untrained dynamical systems composed of a collection of recurrently connected units---by training a readout layer to map the output of the physical system to a target problem \cite{schrauwen2007overview}.

The brain is an excellent example of a self-organizing system; it shows a remarkable capacity for computation with very little energy consumption and no centralized control, and scientists and engineers have long looked to the structure and behavior of the brain for inspiration.
Neurons grown \textit{in vitro} self-organize into networks that show complex patterns of spiking activity, which can be analyzed to gain insight into the network's capacity for information storage and transmission.
This behavior indicates that \textit{in vitro} neuronal networks may serve as a suitable computational reservoir \cite{aaser2017towards} and could also provide insights into the characteristics and dynamics desired for more engineerable substrates.

The aim of the present research project is to construct computational models that are able to reproduce desired behaviors observed in electrophysiological data recorded from engineered neuronal networks.
These models will provide insight into the behavior of the neurons and enable us to reproduce it in other substrates.
The computational capabilities of the models and different physical substrates developed from the models will be explored and their dynamics characterized.
This work is part of a project entitled Self-Organizing Computational substRATES (SOCRATES) \cite{socratesweb},
which aims to take inspiration from the behavior of \textit{in vitro} neuronal networks toward the development of novel self-organizing computing hardwares based in nanomagnetic substrates.

In addition to providing an avenue for the development of novel computational hardwares, the developed models are also expected to provide insight into the functionality of neuronal networks in healthy and perturbed conditions, where typical perturbations include chemical manipulation or electrical stimulation.
The dynamics of perturbed neuronal networks will also be modeled using the developed framework, and the computational capabilities and dynamics of the resulting models will be characterized.
On the basis of this modeling, strategies of interfacing with perturbed networks to recover their dynamics will be explored.
The behavior of perturbed networks and their capacity for recovery will also provide insight into the robustness of the computational capabilities of engineered self-organizing substrates against analogous damage or perturbation.

The remainder of this paper is organized as follows.
Section \ref{background} gives some background on the analysis method used in this work to assess the criticality of \textit{in vitro} neuronal networks toward identifying networks that may be considered well-suited for computation.
Section \ref{methods} describes the methods of preparing the dissociated primary cortical networks evaluated in this study, which were cultured on a microelectrode array (MEA) to enable observation of their electrical behavior, and the analysis methods applied to the electrophysiological data recorded from the networks.
The results from the preliminary analysis of the recorded data are presented in Section \ref{results}.
These results include the observation of the course of maturation of the networks and their response to chemical perturbation to increase inhibition.
The implications of the presented results are discussed in Section \ref{discussion}.
A brief overview of the long-term plan for this research project is then given in Section \ref{plan},
and section \ref{concl} concludes the paper.

\section{Background: Neuronal avalanches} \label{background}
It has been theorized that the brain self-organizes into a critical state to optimize its computational properties; the foundations and theorized functional benefits of this behavior have been reviewed in recent works \cite{shewplenz2013functional, hessegross2014}.
A system in the critical state rests at the boundary between two qualitatively different types of behavior.
In the subcritical phase, a system shows highly ordered behavior characterized as static or oscillating between very few distinct states,
whereas in the supercritical or chaotic phase, the system shows highly unpredictable, essentially random behavior.
Near the transition point between these two regimes, the system is poised to effectively respond to a wide range of inputs as well as store and transmit information, making it ideal in terms of the capacity a system has for computation \cite{langton1990edgeofchaos}.

\begin{figure}
    \centering
    \subfloat[\label{figMEA}]{%
        \includegraphics[width = 0.45\textwidth]{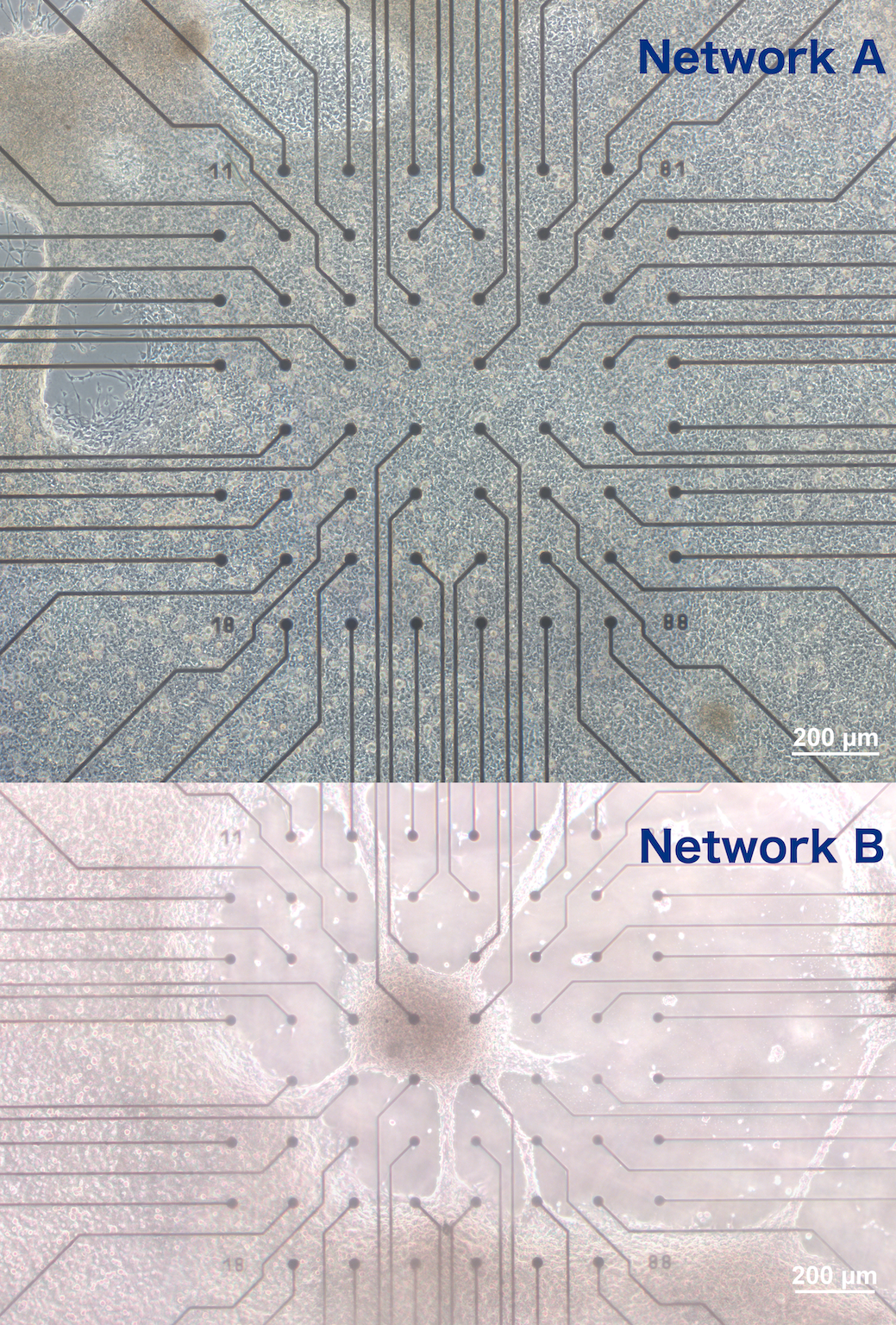}
    }\\ 
    \subfloat[\label{figAvalanche}]{%
        \includegraphics[width = 0.48\textwidth]{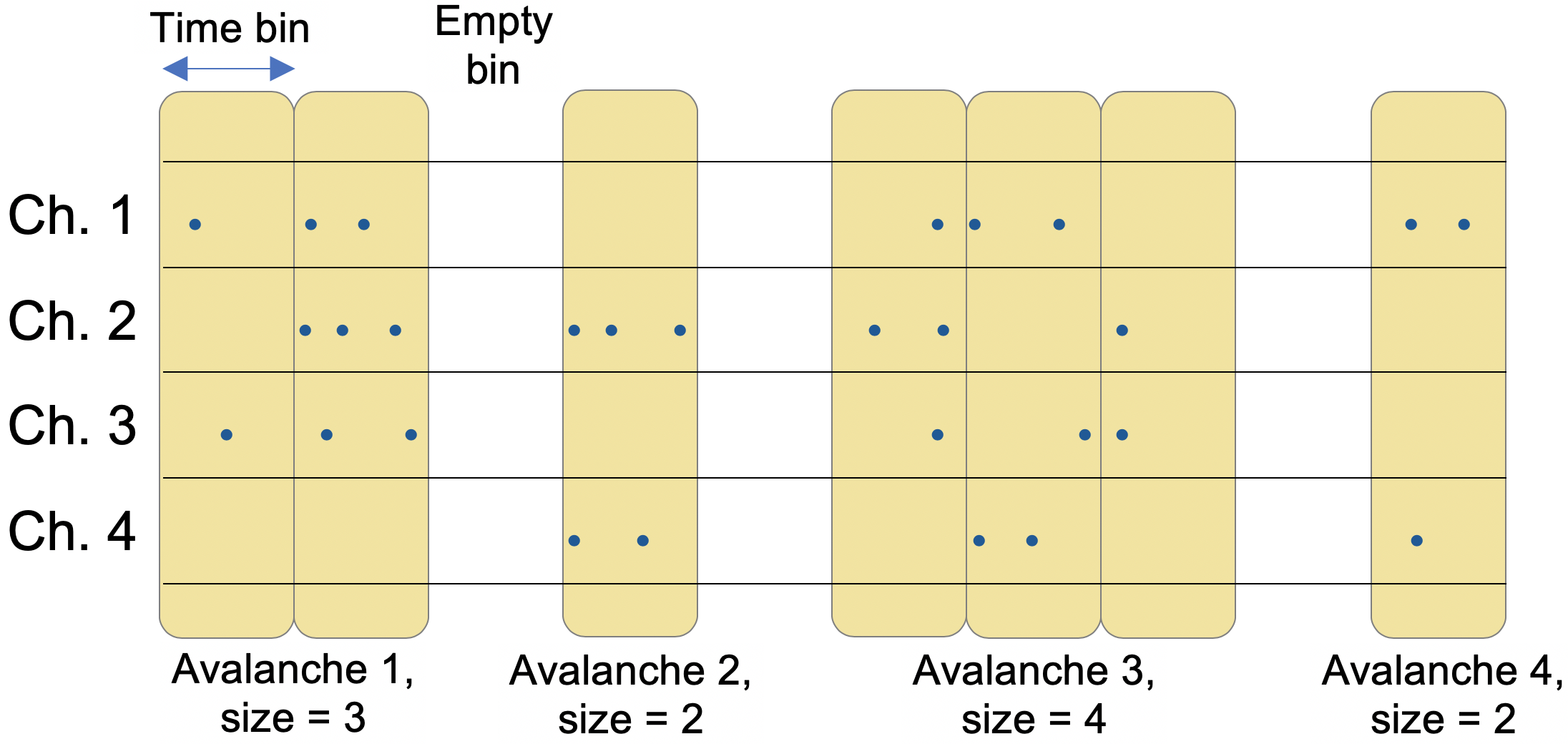}
    }
    \caption{(a) Microscope images of the primary cortical networks analyzed in this study. Network A (shown here at DIV 49) showed activity on all electrodes, whereas Network B (shown here at DIV 51) partially detached from the electrode area of the MEA chip, resulting in approximately 12--15 electrodes of the 60 showing the vast majority of the recorded network activity. (b) Definition of a neuronal avalanche. Each dot represents a spike recorded by one of the electrodes (Ch. 1--4). A time bin is active when it contains at least one spike and empty when there are no spikes. An avalanche is defined as a sequence of consecutive active time bins preceded and followed by empty bins, and the size is the number of electrodes active during the avalanche.}
    \label{fig:avalancheMEA}
\end{figure}

To determine whether a neuronal network is in the critical state, we turn to the scaling behavior of network-wide avalanches of activity.
As first defined by Beggs and Plenz \cite{beggsplenz2003avalanches}, a neuronal avalanche is any number of consecutive time bins in which at least one spike is recorded, bounded before and after by time bins containing no activity, as shown in Fig.~\ref{figAvalanche}.
In their study, Beggs and Plenz \cite{beggsplenz2003avalanches} demonstrated that the probability distributions of the size and duration of neuronal avalanches follow a power law, indicating that the propagation of activity in the cortex is in the critical state \cite{bak1987}.
It has been further demonstrated that criticality is established by a balance between excitation and inhibition and that cortical networks at criticality show 
a larger dynamic range (sensitivity to a wider range of inputs), 
higher information capacity (number of output patterns generated in response to different inputs), 
and greater information transmission (information shared between two recording sites) than networks functioning outside of criticality \cite{shew2009, shew2011}.
Studies on the spontaneous activity of dissociated cortical networks have indicated that these networks tend to self-organize into the critical state over the course of their maturation, though not all networks settle into this state and there is disagreement on when the networks reach criticality \cite{pasquale2008, tetzlaff2010developing, yada2017}.

In the present study, an analysis method based on the size distribution of neuronal avalanches was applied to the analysis of \textit{in vitro} primary cortical neuronal networks (Fig.~\ref{figMEA}).
This method is based on previous analysis performed on cortical networks (e.g., \cite{pasquale2008}).
The aim of this method is to determine whether a given neuronal network is in the critical state, which is presumed to be beneficial for the network in terms of its capacity to store information and perform computation.

As a long-term goal of the present research project, avalanche size distribution analysis is being applied to recordings obtained from different types of \textit{in vitro} neuronal networks (e.g., human induced pluripotent stem cell (iPSC)-derived dopaminergic networks, see \cite{heiney2019iPSC}) to assess the criticality of the networks.
Emerging network dynamics in healthy and perturbed conditions will be studied, characterized, and classified.
On the basis of the preliminary results obtained in this work, we report the development of two cortical networks from day \textit{in vitro} (DIV) 7 to DIV 51 and their response to chemical perturbation on DIV 51.

To the authors' knowledge, this work represents the first time a network has been manipulated into the critical state through chemical perturbation.
Furthermore, this type of analysis has also not yet been applied to the characterization of the neuronal network dynamics in \textit{in vitro} disease models, which is the focus of ongoing and future work.

\section{Methods} \label{methods}
Neuronal networks were prepared using primary cortical neurons.
This section presents the methods for preparing and recording from the neuronal networks, as well as the data analysis methods applied to the recorded data.

\subsection{Preparation and electrophysiology of neuronal networks} \label{culturing}
The neuronal networks assessed here were prepared as follows.
Primary rat cortical neurons (Thermo Fisher) were seeded on a feeder layer of human astrocytes (Gibco, Thermo Fisher) at a density of approximately $1 \times 10^5$ neurons per MEA.
The networks were left to mature for 7 DIV prior to recording.
The spontaneous electrophysiological activity of the network was recorded using a 60-electrode MEA together with the corresponding \textit{in vitro} recording system (MEA2100-System, Multi Channel Systems) and software (Multi Channel Experimenter, Multi Channel Systems).
Recordings were taken every second day for 15 min.
Culture feedings occurred at the start of each week immediately after recording.

$\gamma$-Aminobutyric acid (GABA, Sigma Aldrich) was added to the cortical networks at rising concentrations to disrupt the excitation-to-inhibition ratio by increasing network inhibition.
Prior to perturbation, a recording was taken to establish a baseline.
GABA was then added directly to the culture media in microliter volumes, and recordings were taken immediately after this perturbation.
Different concentrations (10 and 50~\si{\micro}M for Network A and 5, 10, and 25~\si{\micro}M for Network B) were chosen to provide increasing degrees of perturbation; lower concentrations were used for Network B because it showed lower levels of activity.
The 15-min recordings were performed in succession with increasing concentrations.
Following the final perturbation, 90\% of the culture media was replaced to return the cultures to the baseline state. 

\subsection{Avalanche analysis} \label{avalanchemethod}
Avalanches were detected according to the method described by Beggs and Plenz \cite{beggsplenz2003avalanches}.
Spikes were detected using a simple thresholding method \cite{heiney2019spikehunter} based on the standard deviation of the noise of the signal after applying a bandpass filter with a pass band of 300 Hz to 3 kHz\footnote{Code for spike detection is available at \url{https://github.com/SocratesNFR/MCSspikedetection}.}.
The spike detection threshold was set to 6, 7, and 8 standard deviations below the median of the signal for the cortical networks, and the results were compared to assess the effect of the detection threshold on the classification results.

The spikes were binned into time bins equal to the average inter-event interval (IEI), which is the time between events recorded across all electrodes, and avalanches were detected as any number of consecutive active time bins (bins containing at least one spike) bounded before and after by empty time bins (Fig. \ref{figAvalanche}).
The size of an avalanche is defined as the number of electrodes that were active during the avalanche.

A power law was then fitted to the avalanche size distribution data.
This power law takes the form
\begin{equation}
    P(s) \propto s^{-\alpha},
\end{equation}
where $s$ is the avalanche size, $P(s)$ is the probability of an avalanche having size $s$, and $\alpha$ is the exponent of the fitted power law. The exponent has been reported to take a value of $\alpha=1.5$ for \textit{in vitro} neuronal networks \cite{beggsplenz2003avalanches,pasquale2008}.
The fitting was performed using two different fitting methods, nonlinear regression (NLR) and maximum likelihood estimation (MLE), and the results were compared.
The fit was applied over the size range of $s_{min}=2$ to the maximum detected avalanche size $s_{max}$, with a cap at $s_{max}=59$ electrodes.
The goodness of fit was computed following Clauset et al.\ \cite{clauset2009powerlaw}.
Synthetic datasets were generated from the fitted distribution, and their Kolmogorov--Smirnov (KS) distances from the theoretical distribution were compared to the empirical KS distance.
The fitting was rejected if the fraction $p$ of synthetic KS distances that were greater than the empirical KS distance was less than 0.1 ($p<0.1$)\footnote{Code for avalanche detection and goodness of fit evaluation is available at \url{https://github.com/SocratesNFR/avalanche}.}.

\section{Results} \label{results}
The avalanche size distributions of different \textit{in vitro} neuronal networks were observed as the networks matured, and the networks were classified as being critical or not critical at each recording time point based on the fitting results.
In this preliminary work, no rigorous analysis was yet applied to classify networks as sub- or supercritical;
rather, only the goodness of fit of the size distribution to a power law was evaluated to assess whether or not the network was in a critical state during each analyzed recording.
Preliminary classification of non-critical cases as sub- or supercritical was performed by visual inspection alone, where subcritical behavior is characterized by exponential decay and supercritical by a bimodal distribution.

The primary cortical networks evaluated here (Networks A and B; see Fig.~\ref{figMEA}) were observed as they matured from DIV 7 to DIV 51, and then the change in the activity in response to chemical perturbation by GABA on DIV 51 was evaluated.
These networks showed a large amount of activity with high-amplitude spikes, providing richer results than iPSC-derived dopaminergic networks investigated in a previous work \cite{heiney2019iPSC} and allowing for the comparison of the results across multiple detection thresholds (6, 7, and 8 standard deviations below the median) and fitting techniques (NLR and MLE), as described in Section \ref{avalanchemethod}.

This section will first present the observed course of maturation for the two cortical networks based on their avalanche size distributions and then discuss how the addition of GABA affected the criticality of the networks.
In both cases, the avalanche size distributions and fitting results will be compared for the different detection thresholds and fitting methods considered in this work.

\subsection{Spontaneous activity during maturation}
Both networks showed an initial low-activity phase from DIV 7 to DIV 12, in which the mean firing rate averaged across all active electrodes was less than $0.1$~s$^{-1}$ and fewer than approximately 1000 neuronal avalanches were detected within the size range (2--59 electrodes) considered for the power law fitting.
In this time period, there were considered to be too few avalanches for the fitting results to be reliable.
In all subsequent recordings of spontaneous activity (DIVs 14--51), both the NLR and MLE fitting methods produced fits that did not meet the criterion ($p>0.1$) for the networks to be classified as critical, indicating that neither of the networks entered the critical state during the observation period.
All subsequent discussion on classifying the networks as sub- or supercritical over the course of their maturation is thus based solely on visual inspection; as stated previously, the implementation of a rigorous classification method in this regard remains as a task for future work.

\begin{figure}
    \centering
    \subfloat[\label{corticalRasterSub}DIV 21]{%
        \includegraphics[width = 0.225\textwidth]{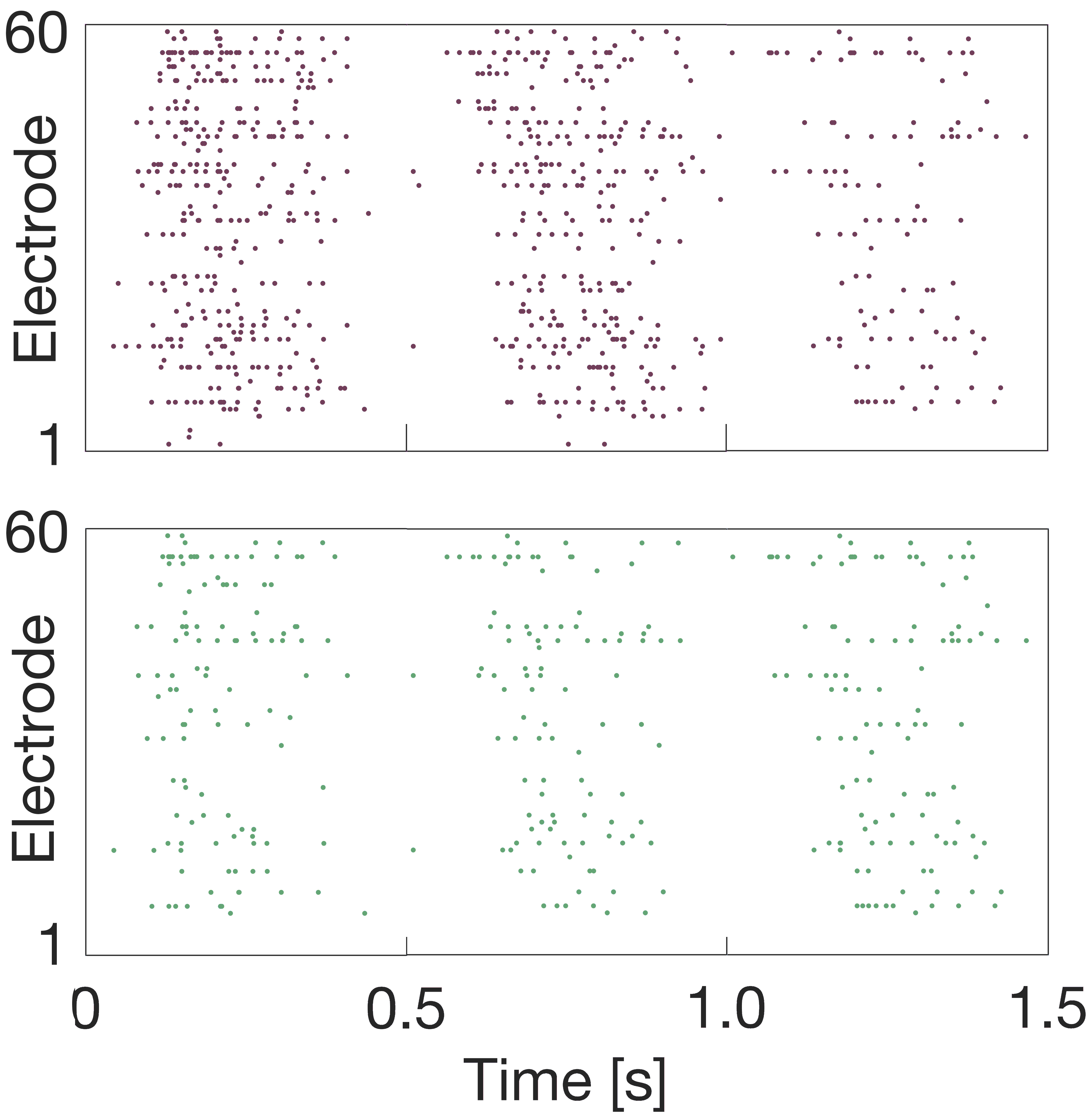}
    }
    \subfloat[\label{corticalPDFsub}DIV 21]{%
        \includegraphics[width = 0.225\textwidth]{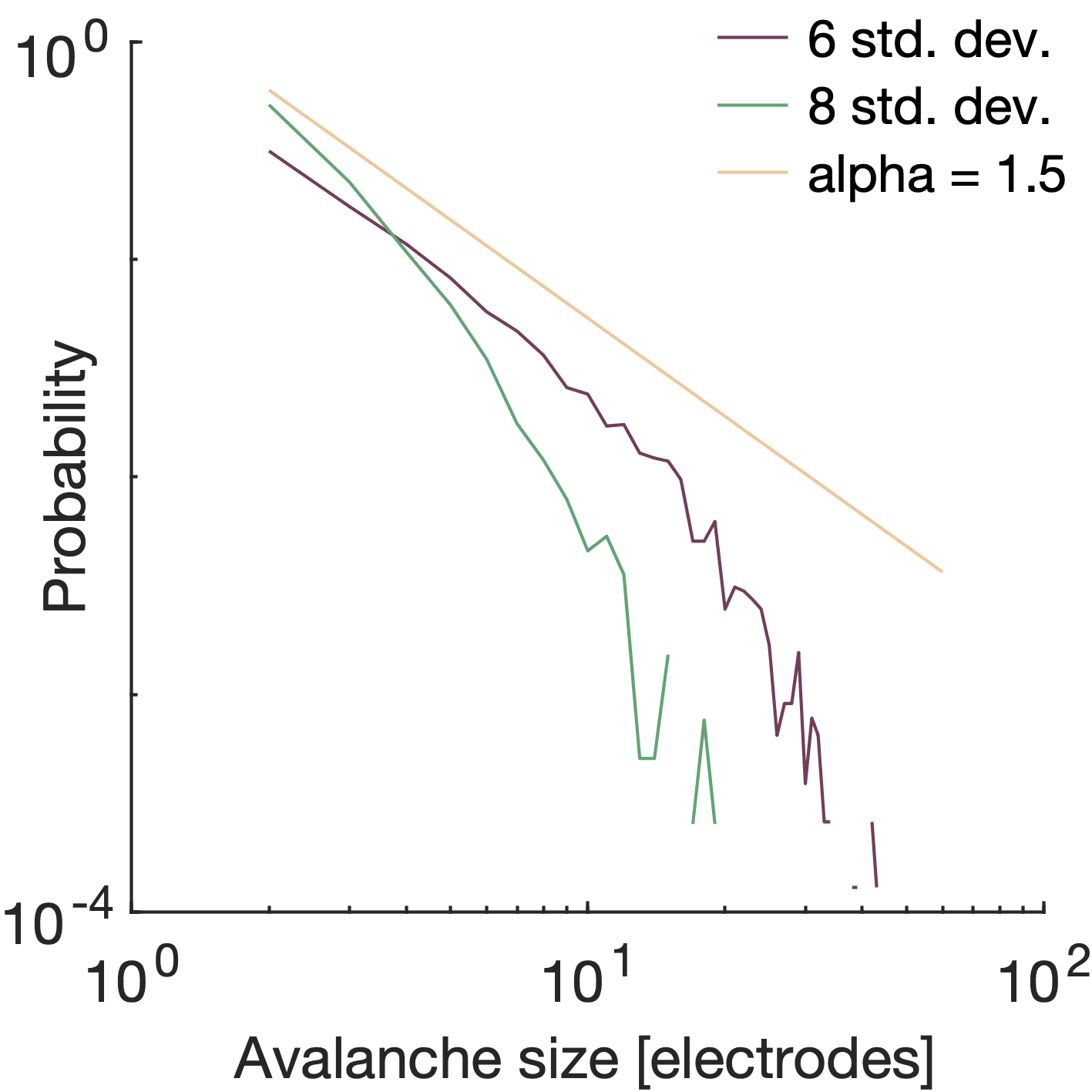}
    }\\
    \subfloat[\label{corticalRasterSuper}DIV 44]{%
        \includegraphics[width = 0.225\textwidth]{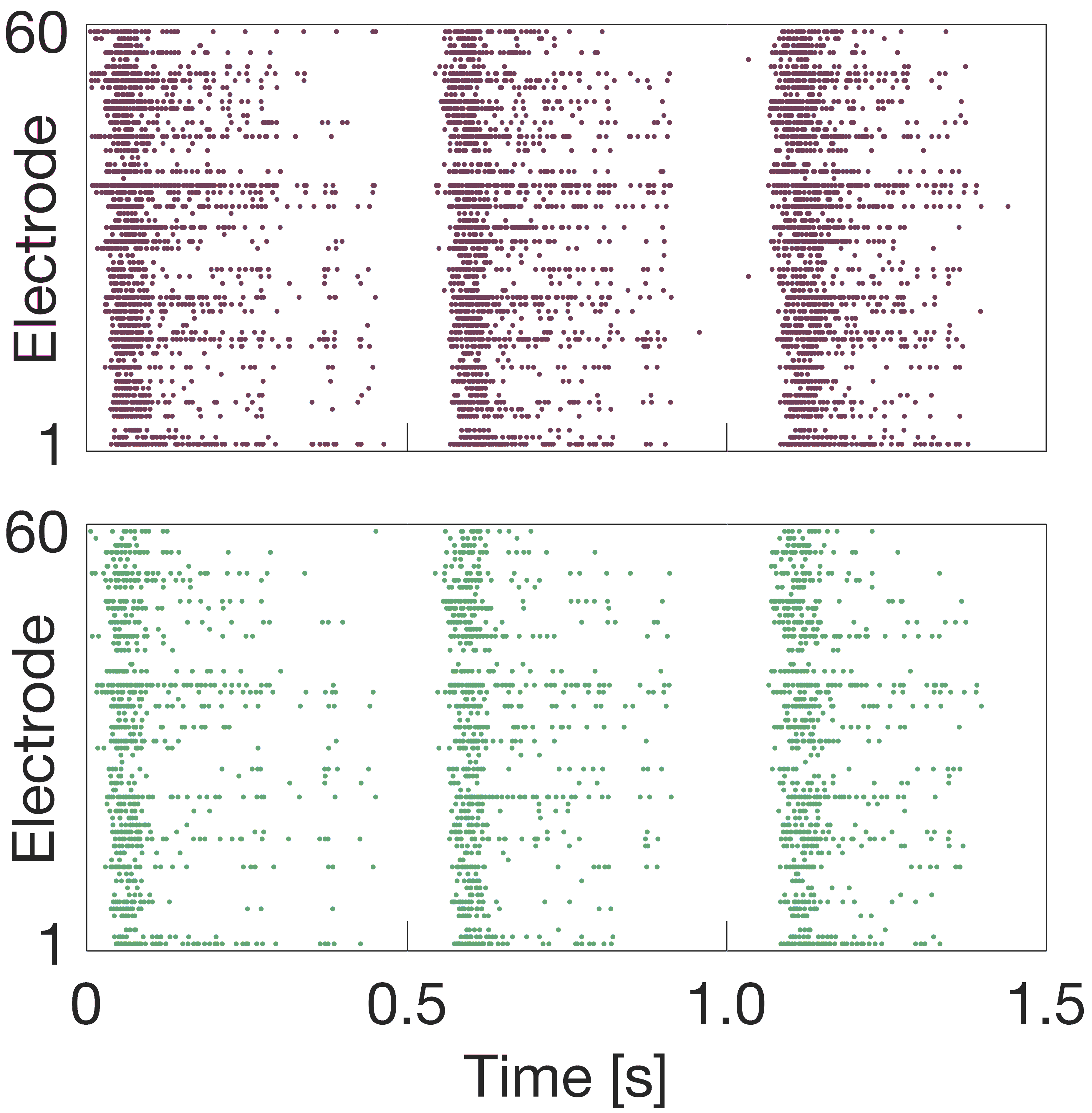}
    }
    \subfloat[\label{corticalPDFsuper}DIV 44]{%
        \includegraphics[width = 0.225\textwidth]{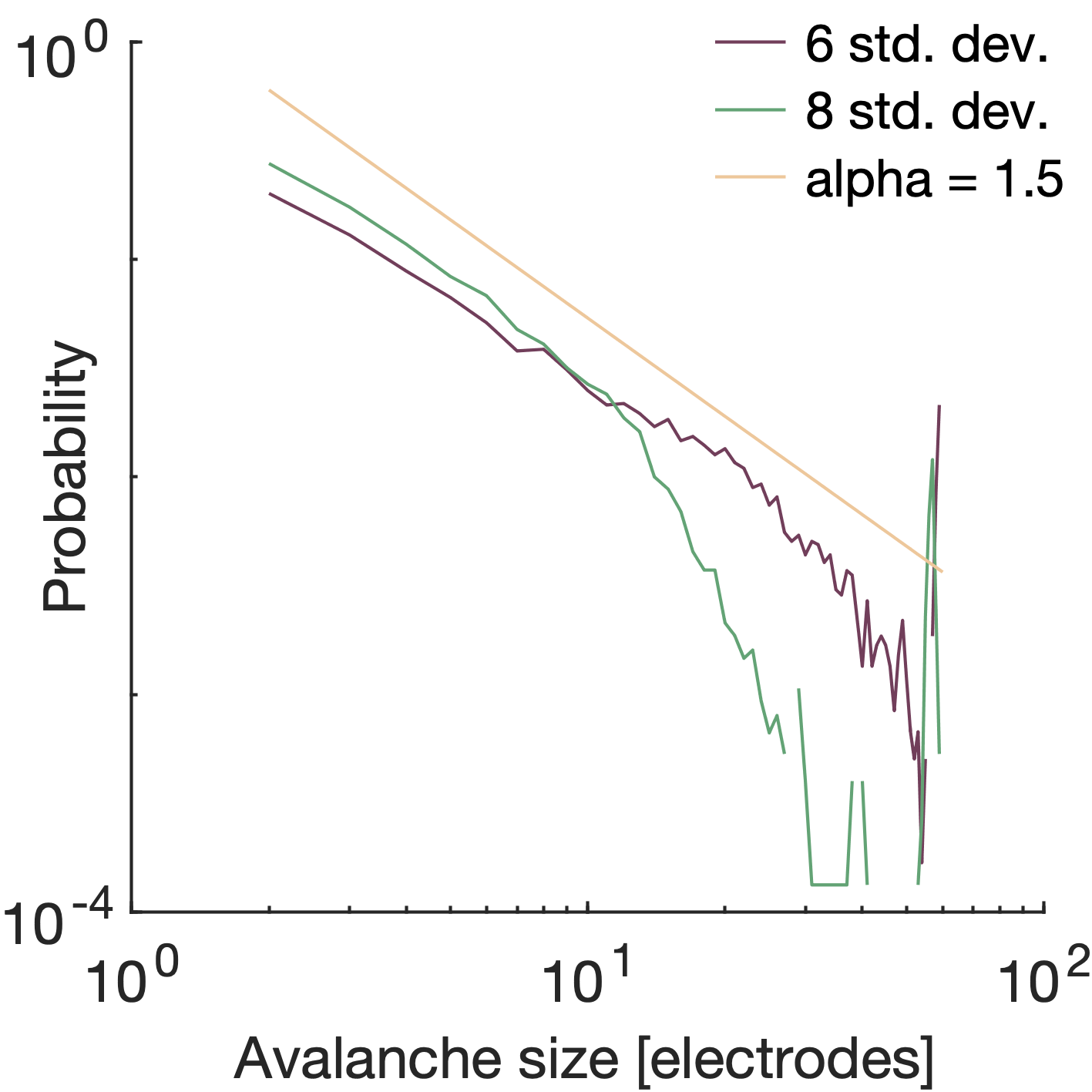}
    }\\
    \subfloat[\label{corticalRasterSurprise}DIV 37]{%
        \includegraphics[width = 0.225\textwidth]{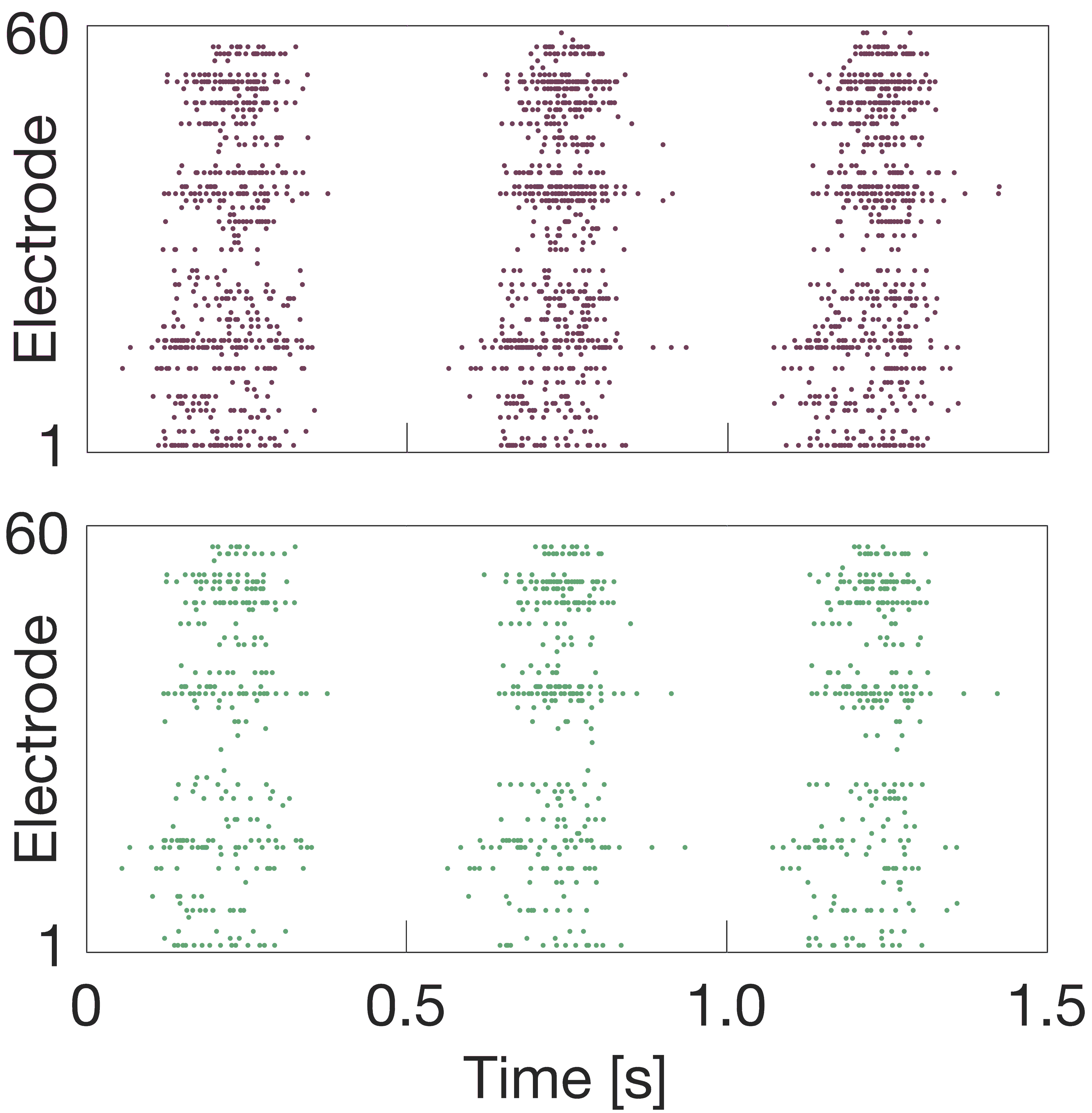}
    }
    \subfloat[\label{corticalPDFsurprise}DIV 37]{%
        \includegraphics[width = 0.225\textwidth]{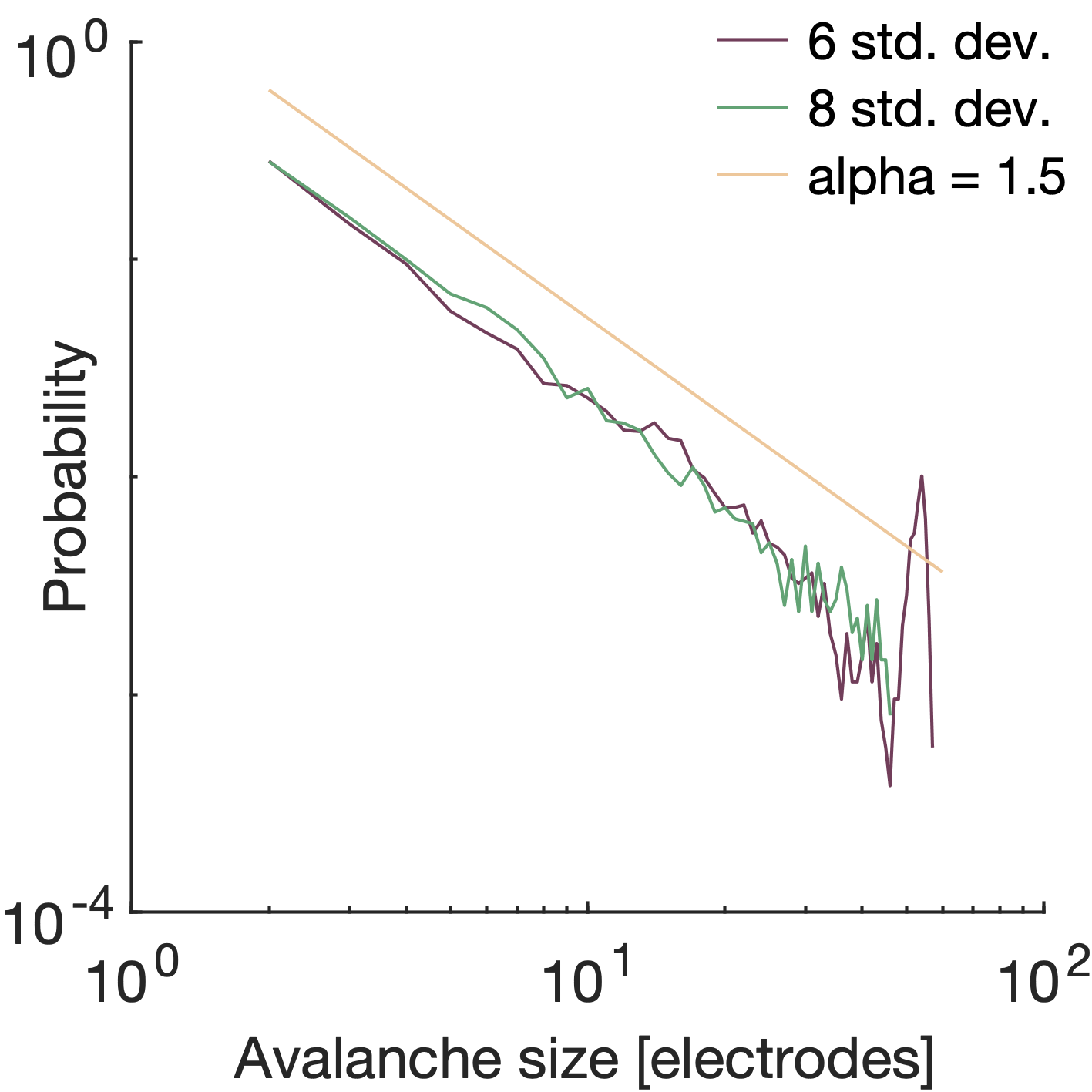}
    }\\
    \caption{Raster plots and size distributions for Network A.
    (a)~DIV 21. Raster plots for detection thresholds of 6 (top) and 8 (bottom) standard deviations. The raster plots show 0.5-s intervals of high network activity with the intervening silent periods eliminated for the sake of visualization.
    (b)~DIV 21. Size distributions for detection thresholds of 6 and 8 standard deviations. A power law function with an exponent of $\alpha=1.5$ is plotted for reference.
    (c,d) Same as (a,b) for DIV 44.
    (e,f) Same as (a,b) for DIV 37.}
    \label{fig:corticalspontaneous}
    \vspace{-5mm}
\end{figure}

Figure \ref{fig:corticalspontaneous} shows a selection of raster plots and avalanche size probability distributions for Network A at three different time points (DIVs 21, 44, and 37).
The raster plots and size distributions are shown for detection thresholds of 6 and 8 standard deviations at each time point.
The raster plots show the first three network bursts in each recording with the periods of relative silence between the bursts removed for the sake of visualization.

The first time point shown (DIV 21; Fig.~\ref{corticalRasterSub} and \ref{corticalPDFsub}) is representative of the type of activity seen from DIV 14 to 21.
During this period, the networks appeared to be in a subcritical state, which is characterized by an exponential size distribution showing a rapid drop in probability as the avalanche size increases.
As shown in the raster plots, this corresponds to loose synchrony and relatively low levels of activity.
The second time point (DIV 44; Fig.~\ref{corticalRasterSuper} and \ref{corticalPDFsuper}) is representative of the type of activity seen from DIV 23 to 51.
During this period, the networks appeared to be in a supercritical state, which is characterized by a bimodal size distribution showing many very large avalanches and few avalanches of intermediate size.
The raster plots demonstrate that this type of activity corresponds to tight synchronicity across most of the electrodes and high activity levels.

The final time point shown in Fig.~\ref{fig:corticalspontaneous} (DIV 37; Fig.~\ref{corticalRasterSurprise} and \ref{corticalPDFsurprise}) is shown to demonstrate the robustness of the classification method based on the goodness of fit measure $p$ against false positives.
The size distribution obtained for a detection threshold of 8 standard deviations appears to follow a power law, showing high linearity in log--log space with a slope corresponding to the expected value of $\alpha=1.5$ (MLE: $\alpha=1.62$, NLR: $\alpha=1.53$).
However, in both cases, the criterion $p>0.1$ was not met (MLE: $p=0.005$, NLR: $p=0.0$).
The analysis results at a threshold of 6 standard deviations more clearly show the deviation of the size distribution from the linear fit, displaying the same type of bimodal distribution seen at DIV 44 (Fig.~\ref{corticalPDFsuper}).
This supports the reliability of the goodness of fit measure.


\subsection{Perturbation to increase inhibition}


\begin{figure}
    \centering
    \subfloat[\label{31528RasterBefore}Network A before perturbation]{%
        \includegraphics[width = 0.45\textwidth]{%
        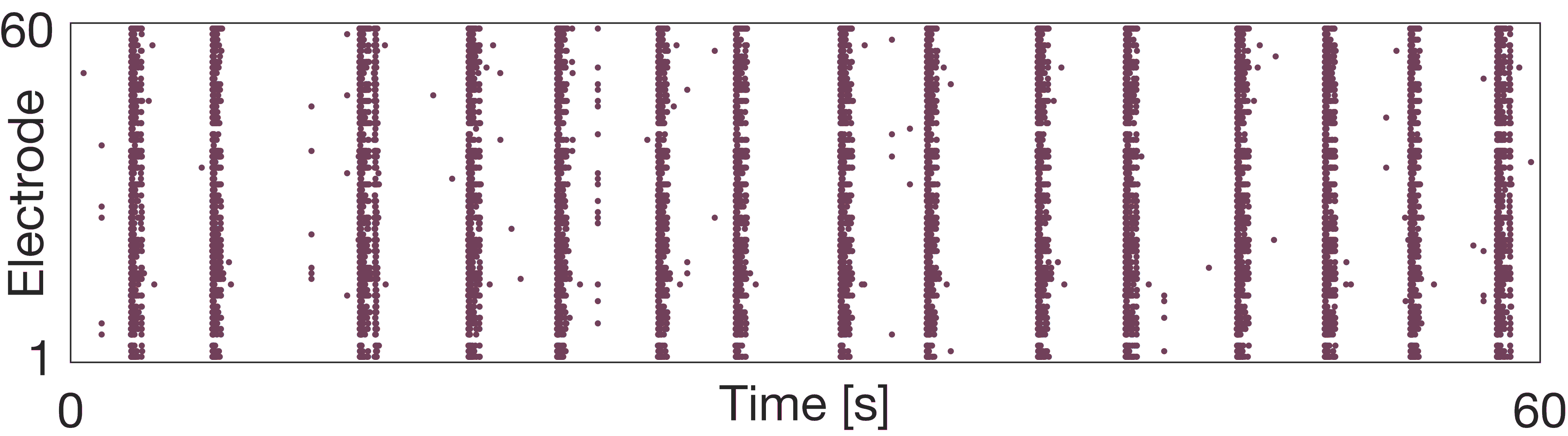}
    }\\
    \vspace{-3mm}
    \subfloat[\label{31528RasterAfter}Network A after perturbation]{%
        \includegraphics[width = 0.45\textwidth]{%
        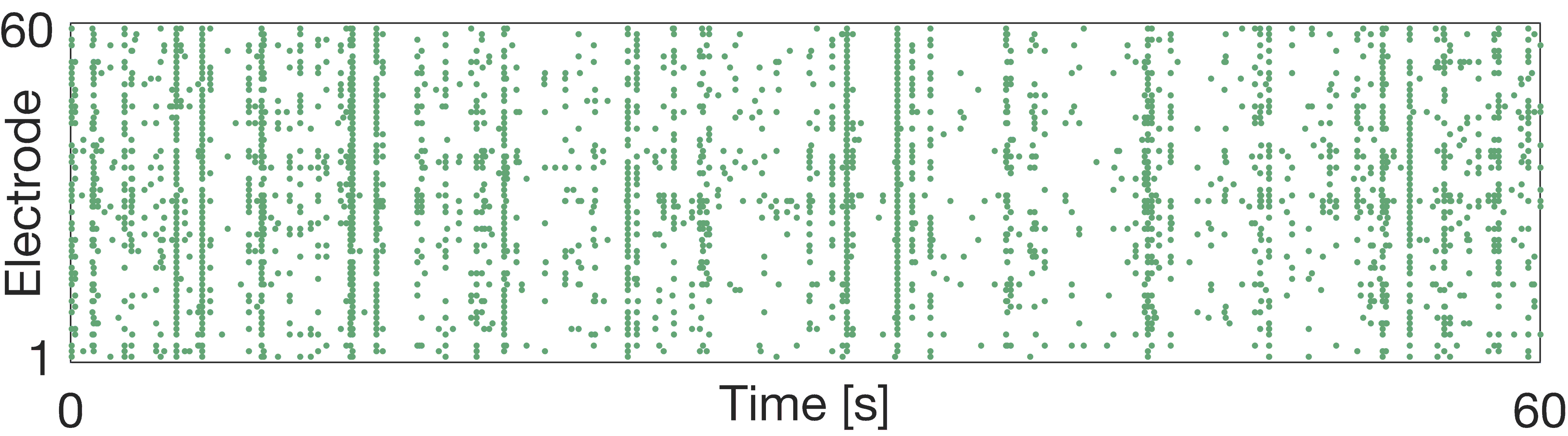}
    }\\
    \subfloat[\label{31528pdfBefore}Network A before perturbation]{%
    \includegraphics[width = 0.225\textwidth]{%
    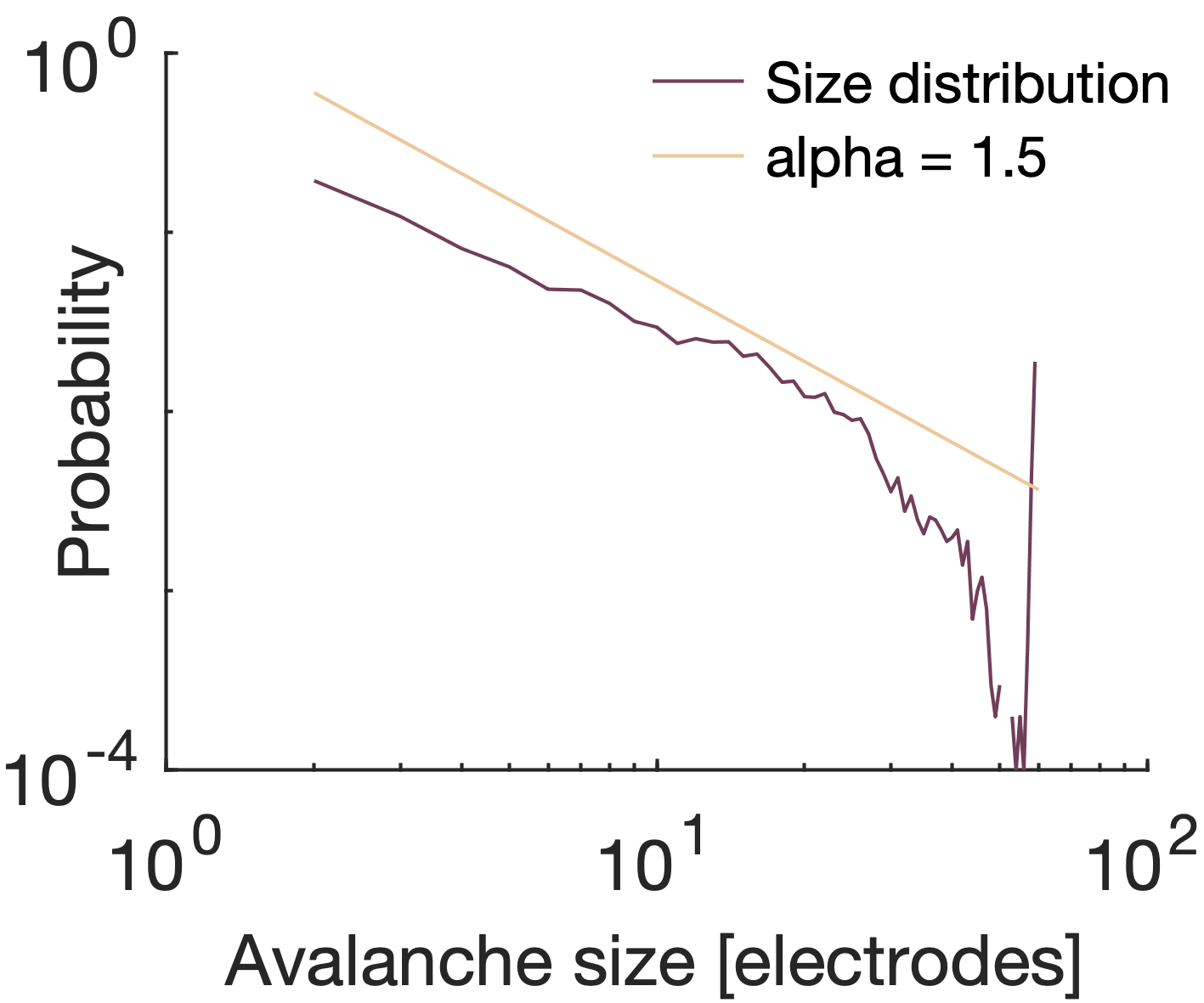}
    }
    \subfloat[\label{31528pdfAfter}Network A after perturbation]{%
    \includegraphics[width = 0.225\textwidth]{%
    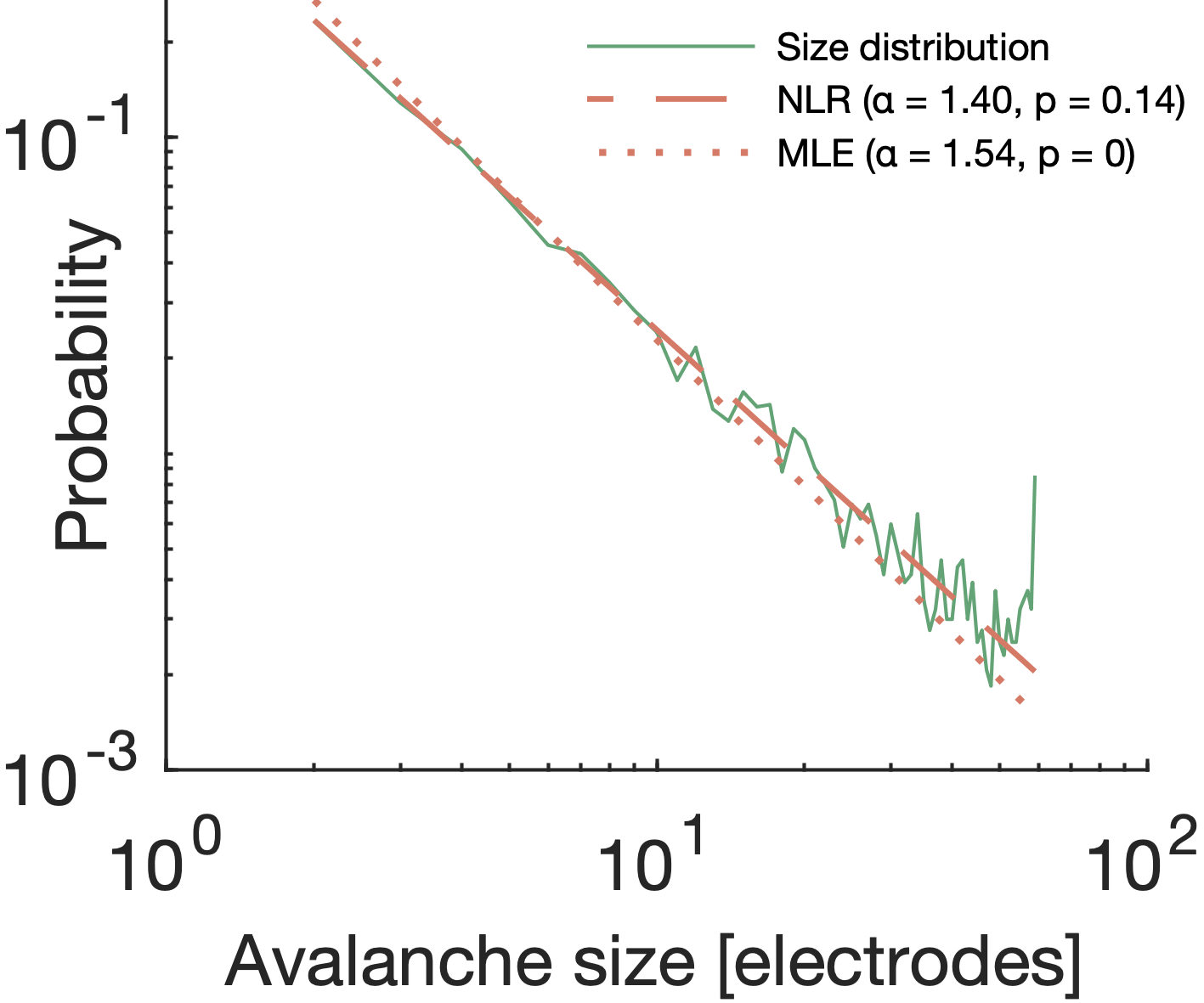}
    }
    
    \subfloat[\label{32428RasterBefore}Network B before perturbation]{%
        \includegraphics[width = 0.45\textwidth]{%
        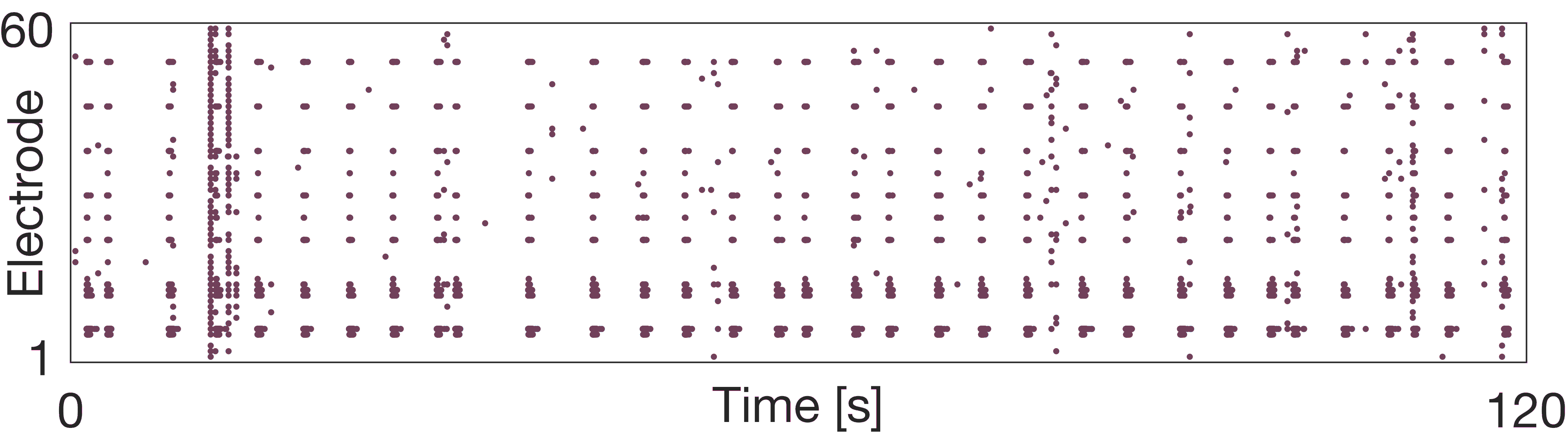}
    }\\
    \vspace{-3mm}
    \subfloat[\label{32428RasterAfter}Network B after perturbation]{%
        \includegraphics[width = 0.45\textwidth]{%
        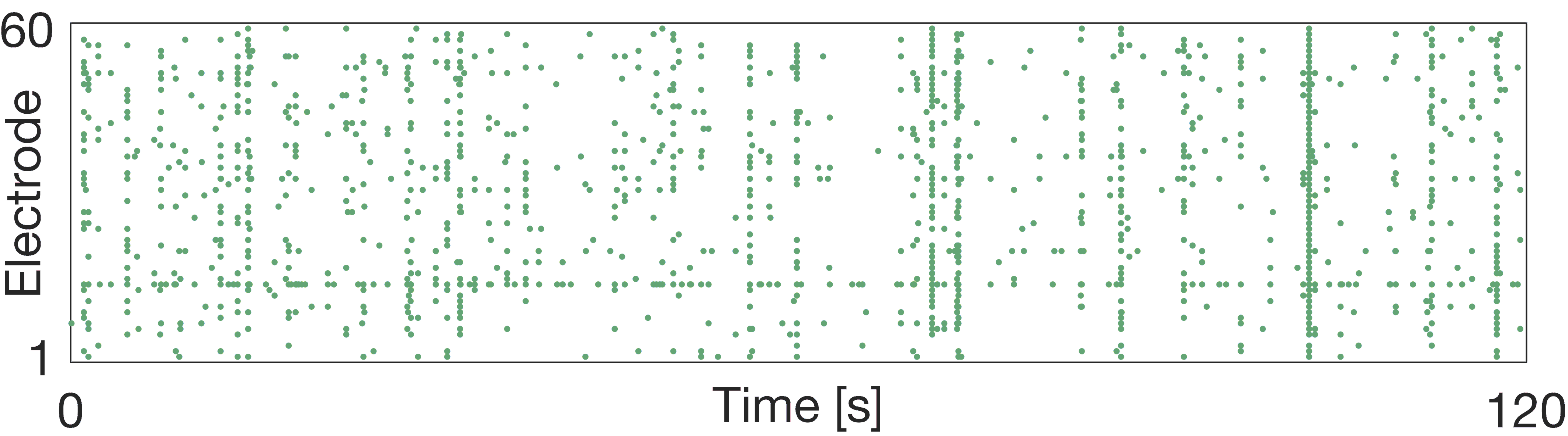}
    }\\
    \subfloat[\label{32428pdfBefore}Network B before perturbation]{%
    \includegraphics[width = 0.225\textwidth]{%
    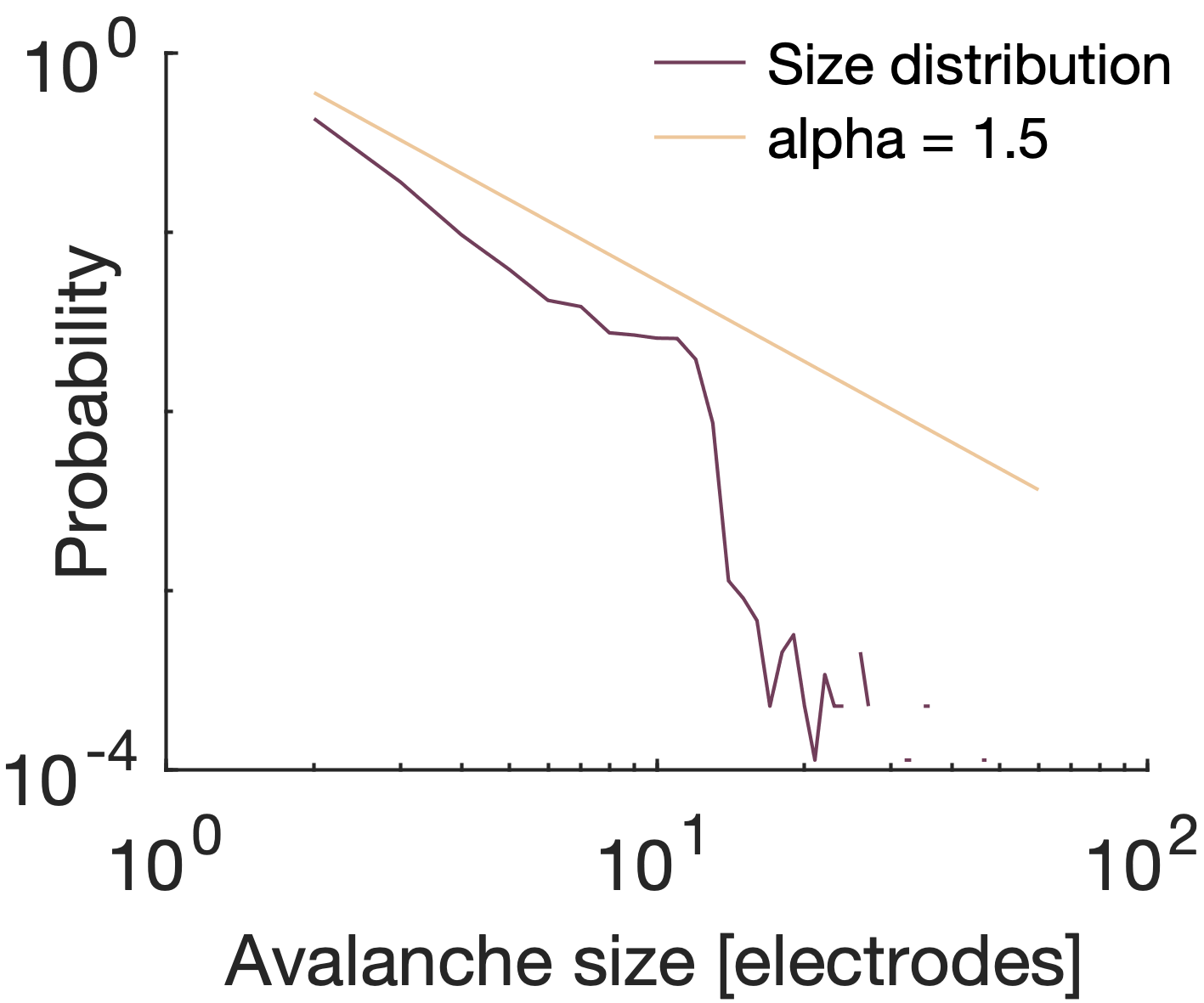}
    }
    \subfloat[\label{32428pdfAfter}Network B after perturbation]{%
    \includegraphics[width = 0.225\textwidth]{%
    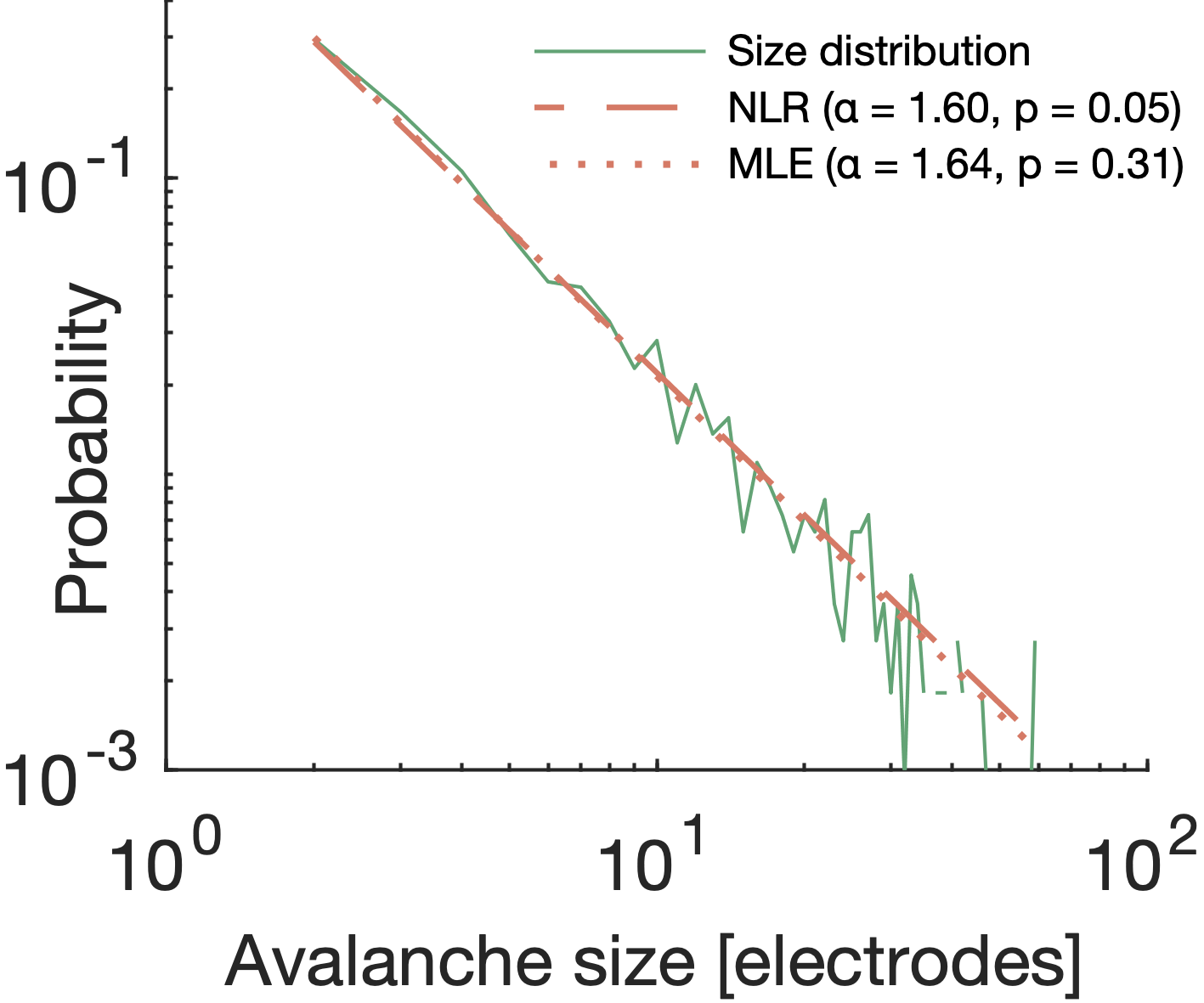}
    }
    \caption{Raster plots and avalanche size distributions before and after chemical perturbation by GABA (DIV 51, threshold of 6 standard deviations).
    (a)~Raster plot of Network A before perturbation.
    (b)~Raster plot of Network A after perturbation.
    (c)~Probability distribution of avalanche sizes for Network A before perturbation. A power law with an exponent of $\alpha=1.5$ is plotted for reference.
    (d)~Probability distribution of avalanches sizes for Network A after perturbation. The fitting results obtained by NLR and MLE are plotted along with their $\alpha$ and $p$ values.
    (e--h) Same as (a--d) for Network B.}
    \label{fig:GABAresults}
\end{figure}

On DIV 51, both networks showed highly synchronized activity and avalanche size distributions with peaks at large avalanche sizes, behavior indicative of a supercritical state.
Figure \ref{31528RasterBefore} and \ref{32428RasterBefore} shows raster plots of the activity of Networks A and B, respectively, on DIV 51 before chemical perturbation, which demonstrate their high level of synchronicity.\footnote{Note that the raster plot for Network A shows 1 min of activity, whereas that for Network B shows 2 min. Because Network B showed much less activity than Network A both before and after perturbation, the type of activity is better captured by showing a larger time window.}
These plots were obtained with a detection threshold of 6 standard deviations.
The corresponding avalanche size distributions are shown in Fig.~\ref{31528pdfBefore} and \ref{32428pdfBefore}, respectively.
As stated previously, Network B detached from some of the electrodes (see Fig.~\ref{figMEA}), causing approximately 12--15 of the electrodes to show the vast majority of the activity, especially during later recordings.
This is apparent in both the raster plot (Fig.~\ref{32428RasterBefore}) and the size distribution (Fig.~\ref{32428pdfBefore}); there is a small peak in the probability of avalanches of size 12, followed by a very sharp drop off toward larger avalanches.

Figure \ref{31528RasterAfter} and \ref{32428RasterAfter} shows the raster plots of Networks A and B, respectively, after the addition of GABA to increase inhibition.
After the addition of GABA, the amount of activity decreased, and the high level of synchronicity was broken.
In contrast to the behavior prior to perturbation, the perturbed networks showed much less periodicity in their activity, with different groupings of electrodes firing within close temporal proximity.
This is reflected in the avalanche size distributions for the two networks shown in Fig.~\ref{31528pdfAfter}, which both appear to follow a power law.
The fitting results obtained using the two fitting methods are also given in the size distribution plots.
For Network A, NLR yielded a good power-law fit, whereas MLE did not; for Network B, the converse was true.

The effect of the detection threshold and fitting method was further evaluated by assessing the agreement among the fitting results for the same recording under different evaluation parameters.
The fitting results for the two networks after perturbation on DIV 51 are given in Table \ref{tab:fittingresults}, with cases yielding a good fit (indicating the network is in the critical state) reported in bold.
In cases where the number $N$ of detected avalanches was insufficient to produce a reliable fitting ($N\lesssim1000$), the results are shown in italics.
For Network A, the NLR fitting results indicate the network is in the critical state, with reliable results produced for thresholds of 6 and 7 standard deviations.
For Network B, only a threshold of 6 standard deviations yielded enough activity for a reliable fitting, and the MLE results indicate the network is in the critical state.

\begin{table}[]
    \centering
    \caption{Power law fitting results for the two networks after chemical perturbation on DIV 51. The results are given for different spike detection thresholds and fitting methods. Bold text indicates a good fit ($p>0.1$). Italicized results indicate that the number of avalanches $N$ was considered insufficient for reliable fitting results ($N\lesssim1000$).}
    \label{tab:fittingresults}
    \begin{tabular}[c]{c c c c c}
        \hline
         Network & Threshold & Fitting method & $\alpha$ & $p$ \\
         \hline
         \hline
         \multirow{6}{*}{\shortstack{A \\ \\ (50~\si{\micro}M \\ \\ GABA)}} %
           & \multirow{2}{*}{8 st.\ dev.} %
                        & NLR & \textit{1.32} & \textit{0.28} \\
           &            & MLE & \textit{1.54} & \textit{0.02} \\
           & \multirow{2}{*}{7 st.\ dev.} %
                        & NLR & \textbf{1.42} & \textbf{0.72} \\
           &            & MLE & 1.54 & 0.00 \\
           & \multirow{2}{*}{6 st.\ dev.} %
                        & NLR & \textbf{1.40} & \textbf{0.14} \\
           &            & MLE & 1.54 & 0.00 \\
         \hline
         \multirow{6}{*}{\shortstack{B \\ \\ (25~\si{\micro}M \\ \\ GABA)}} %
           & \multirow{2}{*}{8 st.\ dev.} %
                        & NLR & \textit{1.95} & \textit{0.54} \\
           &            & MLE & \textit{1.67} & \textit{0.48} \\
           & \multirow{2}{*}{7 st.\ dev.} %
                        & NLR & \textit{1.72} & \textit{0.16} \\
           &            & MLE & \textit{1.69} & \textit{0.06} \\
           & \multirow{2}{*}{6 st.\ dev.} %
                        & NLR & 1.60 & 0.05 \\
           &            & MLE & \textbf{1.64} & \textbf{0.31} \\
         \hline
    \end{tabular}
\end{table}

\section{Discussion} \label{discussion}

The two cortical networks observed in this study never self-organized into the critical state; rather, they appeared to enter into a subcritical state after an initial low-activity phase and then settled into a stable highly synchronized supercritical state.
Pasquale et al.~\cite{pasquale2008} have demonstrated that not all dissociated cortical networks will reach the critical state during their normal course of maturation.
They also reported that cultures tend to fall into a single preferred state (subcritical, supercritical, or critical) by around DIV 21 and tend not to diverge from that state.
However, other previous studies have demonstrated different courses of maturation for cortical networks.
For example, Yada et al.~\cite{yada2017} have indicated that cortical networks undergo early development from subcritical to supercritical before finally reaching a critical state after DIV 10.
In contrast to this, Tetzlaff et al.~\cite{tetzlaff2010developing} have described activity traversing from supercritical to subcritical and finally reaching criticality around DIV 58 ($\pm$~20 days).
It is possible that the present networks may have reached criticality if left to mature further; however, it seems more likely that they would have remained in the supercritical state, as would be predicted by the type of trends observed by Pasquale et al.~\cite{pasquale2008}.
It will be necessary to perform more rigorous fittings of exponential curves to the size distribution to definitively demonstrate that the cultures showed subcritical behavior.
This remains a task for future work.

An important achievement in this work was the successful perturbation of the \textit{in vitro} networks from supercritical to critical.
The networks showed increasingly synchronized activity accompanied by heightened activity levels as they matured from DIV 23 onward.
A balanced excitation-to-inhibition ratio is widely considered to play an important role in enabling the emergence of critical behaviors, and previous studies have demonstrated the importance of this balance both through modeling (e.g., \cite{pasquale2008, massobrio2015}) and experimentally (e.g., \cite{shew2009, shew2011}).
On the basis of this evidence, the networks were perturbed at DIV 51, when they showed highly synchronized supercritical behavior, by adding GABA;
the aim of this perturbation was to artificially balance the excitation-to-inhibition ratio by increasing inhibition, thereby pushing the networks into the critical regime.
The results indicate that the addition of GABA (50~\si{\micro}M to Network A and 25~\si{\micro}M to Network B) successfully broke the high level of synchronicity seen in the networks and brought them into the critical state (see Fig.~\ref{fig:GABAresults}).
As stated in Section \ref{methods}, the GABA concentration was incrementally raised for each network (10 and 50~\si{\micro}M for Network A; 5, 10, and 25~\si{\micro}M for Network B).
In both cases, the network only reached criticality at the highest considered molarity; however, in future work, smaller increments will be considered, as GABA also has the undesirable effect of reducing activity.

The analysis of both the spontaneous and perturbed activity also demonstrated that it is important to consider different spike detection thresholds and fitting methods when conducting the analysis.
The preliminary results reported here show that the goodness of fit measure \cite{clauset2009powerlaw} was able to robustly identify non-critical behavior even when the size distribution showed apparent linearity in log--log space and that lowering the threshold in this case supported the identification of supercritical activity (see Fig.~\ref{corticalPDFsurprise}).
When GABA was added, there was insufficient activity at certain thresholds to achieve a reliable fitting (see Table \ref{tab:fittingresults}), indicating the importance of an appropriate threshold.
Furthermore, although the results appear to demonstrate critical behavior in Network B with a threshold of 6 standard deviations, it is possible that this threshold was too low and may have yielded false positives, as it is unlikely that there was actual activity on the electrodes where neurons had not adhered (see Fig.~\ref{figMEA}).
In the current experimental setup, a threshold of 7 standard deviations seems to have been most appropriate to balance these considerations.
The results also indicate that more work is needed to determine whether NLR or MLE is best suited to yielding a good power law fit, as the two fittings did not produce consistent values of $\alpha$ or classification results.


\section{Plan for future research} \label{plan}
This work represents a preliminary step in a larger research project, which will be described briefly here.
The plan for this research project is divided into four stages.
In the first stage, a data analysis framework will be developed, with the avalanche analysis method described here constituting a crucial part of this framework.
The framework involves methods of extracting meaningful features from electrophysiological data recorded from \textit{in vitro} neuronal networks.
Such features include conventional parameters considered in electrophysiological data analysis, such as the mean firing rate, as well as more complex measures, such as entropy and measures of connectivity.
The connectivity of the engineered networks will also be modeled using graph theory approaches.
The avalanche method presented here represents a useful tool for classifying networks as critical or non-critical.
Other methods of classification and clustering of networks will also be explored.

The second stage of the project involves the construction of computing models, such as cellular automata (CAs), random Boolean networks (RBNs), and recurrent neural networks (RNNs), that show behavior similar to that of the neuronal networks \cite{nichele2017deeplearningCA, nichele2017reservoirCA}.
The data analysis framework developed in the first phase will be used as a method of capturing the target behavior to be reproduced in the models,
and this framework will be continually refined as we improve our understanding of the important aspects of neuronal behavior that contribute to their computational capabilities.
These computing models are developed using evolutionary algorithms with appropriate fitness functions defined on the basis of the target behavior.
Important features of the models,
such as their input and output mappings and number of states,
will be explored,
and the dynamics of the models will be characterized.

The third phase involves the use of the developed models and the \textit{in vitro} neuronal networks as reservoirs to perform computational and classification tasks as a proof-of-concept using reservoir computing.
The models from the second stage will be refined based on their performance as computing reservoirs.

The final stage consists of the exploration of the application of the models developed in the second stage to the study of engineered neuronal networks under perturbed conditions mimicking pathologies related to the central nervous system (CNS).
Networks that have had their synaptic function perturbed will be modeled and analyzed using the developed methods to characterize how their behavior differs from that of unperturbed networks.
Methods of interfacing with the perturbed networks to restore their dynamics to the unperturbed state will then be explored.

\section{Conclusion} \label{concl}
The aim of this research project is to extract meaningful behaviors and features from electrophysiological data recorded from \textit{in vitro} neuronal networks and construct models that reproduce these behaviors toward the eventual realization of novel computing substrates based in nanomagnetic materials.
This paper reported the application of an avalanche size distribution analysis to electrophysiological data, representing a first step in the development of an analytical framework to extract target behaviors from such data.
The results indicate that the avalanche analysis method applied here is able to successfully classify networks as critical or not critical and that chemical perturbation is a feasible method of inducing criticality when a network is in the supercritical state.
In future work, more networks will be analyzed with the hopes of achieving self-organized criticality in some networks, and further investigations of the effects of chemical perturbation on both critical and supercritical networks will be conducted.

With this type of analysis, it can be determined if a network is in a critical state, which gives an indication of its suitability for use in computational tasks.
Furthermore, the results presented here indicate that it is possible to bring cortical networks from the supercritical state to criticality by increasing inhibition in the networks.
This would be a useful approach to manipulating networks to make them viable for performing computational tasks.
In addition to the computational applications of this analysis, it is also expected to be useful in distinguishing healthy and perturbed networks and to provide insight into how different diseases affect neuronal connectivity and communication, which will be the target of future work.

\section*{Acknowledgements}
This work was conducted as part of the SOCRATES project, which is partially funded by the Norwegian Research Council (NFR) through their IKTPLUSS research and innovation action on information and communication technologies under the project agreement 270961.

\bibliographystyle{IEEEtran}
\bibliography{bib}

\begin{thebibliography}{10}
\providecommand{\url}[1]{#1}
\csname url@samestyle\endcsname
\providecommand{\newblock}{\relax}
\providecommand{\bibinfo}[2]{#2}
\providecommand{\BIBentrySTDinterwordspacing}{\spaceskip=0pt\relax}
\providecommand{\BIBentryALTinterwordstretchfactor}{4}
\providecommand{\BIBentryALTinterwordspacing}{\spaceskip=\fontdimen2\font plus
\BIBentryALTinterwordstretchfactor\fontdimen3\font minus
  \fontdimen4\font\relax}
\providecommand{\BIBforeignlanguage}[2]{{%
\expandafter\ifx\csname l@#1\endcsname\relax
\typeout{** WARNING: IEEEtran.bst: No hyphenation pattern has been}%
\typeout{** loaded for the language `#1'. Using the pattern for}%
\typeout{** the default language instead.}%
\else
\language=\csname l@#1\endcsname
\fi
#2}}
\providecommand{\BIBdecl}{\relax}
\BIBdecl

\bibitem{langton1990edgeofchaos}
C.~G. Langton, ``Computation at the edge of chaos: Phase transitions and
  emergent computation,'' \emph{Physica D}, vol.~40, pp. 12--37, 1990.

\bibitem{broersma2017computational}
H.~Broersma, J.~F. Miller, and S.~Nichele, ``Computational matter: Evolving
  computational functions in nanoscale materials,'' in \emph{Advances in
  Unconventional Computing}.\hskip 1em plus 0.5em minus 0.4em\relax Springer,
  2017, pp. 397--428.

\bibitem{konkoli2018reservoir}
Z.~Konkoli, S.~Nichele, M.~Dale, and S.~Stepney, ``Reservoir computing with
  computational matter,'' in \emph{Computational Matter}.\hskip 1em plus 0.5em
  minus 0.4em\relax Springer, 2018, pp. 269--293.

\bibitem{jensen2018spinice}
J.~H. Jensen, E.~Folven, and G.~Tufte, ``Computation in artificial spin ice,''
  \emph{The 2018 Conference on Artificial Life: A Hybrid of the European
  Conference on Artificial Life (ECAL) and the International Conference on the
  Synthesis and Simulation of Living Systems (ALIFE)}, no.~30, pp. 15--22,
  2018.

\bibitem{heylighen1999}
F.~Heylighen, ``The science of self-organization and adaptivity,'' in \emph{in:
  Knowledge Management, Organizational Intelligence and Learning, and
  Complexity, in: The Encyclopedia of Life Support Systems, EOLSS}.\hskip 1em
  plus 0.5em minus 0.4em\relax Publishers Co. Ltd, 1999, pp. 253--280.

\bibitem{doursat2013morphogenetic}
R.~Doursat, H.~Sayama, and O.~Michel, ``A review of morphogenetic
  engineering,'' \emph{Natural Computing}, vol.~12, pp. 517--535, 2013.

\bibitem{schrauwen2007overview}
B.~Schrauwen, D.~Verstraeten, and J.~Van~Campenhout, ``An overview of reservoir
  computing: Theory, applications and implementations,'' in \emph{Proceedings
  of the 15th European Symposium on Artificial Neural Networks. p. 471-482
  2007}, 2007, pp. 471--482.

\bibitem{aaser2017towards}
P.~Aaser, M.~Knudsen, O.~Huse~Ramstad, R.~van~de Wijdeven, S.~Nichele,
  I.~Sandvig, G.~Tufte, U.~S. Bauer, {\O}.~Halaas, S.~Hendseth \emph{et~al.},
  ``Towards making a cyborg: A closed-loop reservoir-neuro system,'' in
  \emph{Proceedings of the European Conference on Artificial Life 2017}.\hskip
  1em plus 0.5em minus 0.4em\relax MIT Press, 2017.

\bibitem{socratesweb}
\BIBentryALTinterwordspacing
(2018) {SOCRATES}: Self-organizing computational substrates. [Online].
  Available: \url{https://www.ntnu.edu/socrates}
\BIBentrySTDinterwordspacing

\bibitem{shewplenz2013functional}
W.~L. Shew and D.~Plenz, ``The functional benefits of criticality in the
  cortex,'' \emph{The Neuroscientist}, vol.~19, no.~1, pp. 88--100, 2013.

\bibitem{hessegross2014}
J.~Hesse and T.~Gross, ``Self-organized criticality as a fundamental propertie
  of neural systems,'' \emph{Frontiers in Systems Neuroscience}, vol.~8, 2014.

\bibitem{beggsplenz2003avalanches}
J.~M. Beggs and D.~Plenz, ``Neuronal avalanches in neocortical circuits,''
  \emph{The Journal of Neuroscience}, vol.~23, no.~35, pp. 11\,167--11\,177,
  2003.

\bibitem{bak1987}
P.~Bak, C.~Tang, and K.~Wiesenfeld, ``Self-organized criticality: An
  explanation of the 1/$f$ noise,'' \emph{Physical Review Letters}, vol.~59,
  pp. 381--384, Jul 1987.

\bibitem{shew2009}
W.~L. Shew, H.~Yang, T.~Petermann, R.~Roy, and D.~Plenz, ``Neuronal avalanches
  imply maximum dynamic range in cortical networks at criticality,''
  \emph{Journal of Neuroscience}, vol.~29, no.~49, pp. 15\,595--15\,600, 2009.

\bibitem{shew2011}
W.~L. Shew, H.~Yang, S.~Yu, R.~Roy, and D.~Plenz, ``Information capacity and
  transmission are maximized in balanced cortical networks with neuronal
  avalanches,'' \emph{Journal of Neuroscience}, vol.~31, no.~1, pp. 55--63,
  2011.

\bibitem{pasquale2008}
V.~Pasquale, P.~Massobrio, L.~L. Bologna, M.~Chiappalone, and S.~Martinoia,
  ``Self-organization and neuronal avalanches in networks of dissociated
  cortical neurons,'' \emph{Neuroscience}, vol. 153, pp. 1354--1369, 2008.

\bibitem{tetzlaff2010developing}
C.~Tetzlaff, S.~Okujeni, U.~Egert, W{\:o}rg{\:o}tter, and M.~Butz,
  ``Self-organized criticality in developing neuronal networks,'' \emph{PLoS
  Computational Biology}, vol.~6, no.~12, 2010.

\bibitem{yada2017}
Y.~Yada, T.~Mita, A.~Sanada, R.~Yano, D.~J. Bakkum, A.~Hierlemann, and
  H.~Takahashi, ``Development of neural population activity toward
  self-organized criticality,'' \emph{Neuroscience}, pp. 55--65, 2017.

\bibitem{heiney2019iPSC}
K.~{Heiney}, V.~{Devold Valderhaug}, I.~{Sand vig}, A.~{Sandvig}, G.~{Tufte},
  H.~{Lewi Hammer}, and S.~{Nichele}, ``{Evaluation of the criticality of in
  vitro neuronal networks: Toward an assessment of computational capacity},''
  \emph{arXiv e-prints}, p. arXiv:1907.02351, Jul 2019.

\bibitem{heiney2019spikehunter}
\BIBentryALTinterwordspacing
K.~Heiney, J.~Mateus, C.~D.~F. Lopes, E.~Neto, M.~Lamghari, and P.~Aguiar,
  ``{\si{\micro}SpikeHunter}: An advanced computational tool for the analysis
  of neuronal communication and action potential propagation in microfluidic
  platforms,'' \emph{Scientific Reports}, vol.~9, no.~1, p. 5777, 2019.
  [Online]. Available: \url{https://doi.org/10.1038/s41598-019-42148-3}
\BIBentrySTDinterwordspacing

\bibitem{clauset2009powerlaw}
A.~Clauset, C.~R. Shalizi, and M.~E.~J. Newman, ``Power-law distributions in
  empirical data,'' \emph{SIAM Review}, vol.~51, no.~4, pp. 661--703, 2009.

\bibitem{massobrio2015}
P.~Massobrio, V.~Pasquale, and S.~Martinoia, ``Self-organized criticality in
  cortical assemblies occurs in concurrent scale-free and small-world
  networks,'' \emph{Scientific Reports}, 2015.

\bibitem{nichele2017deeplearningCA}
S.~Nichele and A.~Molund, ``Deep learning with cellular automaton-based
  reservoir computing,'' \emph{Complex Systems}, vol.~26, 2017.

\bibitem{nichele2017reservoirCA}
S.~Nichele and M.~S. Gundersen, ``Reservoir computing using nonuniform binary
  cellular automata,'' \emph{Complex Systems}, vol.~26, 2017.

\end{thebibliography}

\end{document}